# Fast Black-Box Quantum State Preparation Based on Linear Combination of Unitaries

Shengbin Wang[1,*], Zhimin Wang[1,*], Guolong Cui[1], Shangshang Shi[1], Ruimin Shang[1], Lixin Fan[1], Wendong Li[1], Zhiqiang Wei[1,2,†] and Yongjian Gu[1,†]

[1] College of Information Science and Engineering, Ocean University of China, Qingdao 266100, China
[2] High Performance Computing Center, Pilot National Laboratory for Marine Science and Technology (Qingdao), Qingdao 266100, China
[*] These authors contribute equally to this work.
[†] Correspondence author, e-mail: guyj@ouc.edu.cn; weizhiqiang@ouc.edu.cn

**Abstract**
Black-box quantum state preparation is a fundamental primitive in quantum algorithms. Starting from Grover, a series of techniques have been devised to reduce the complexity. In this work, we propose to perform black-box state preparation using the technique of linear combination of unitaries (LCU). We provide two algorithms based on a different structure of LCU. Our algorithms improve upon the existed best results by reducing the required additional qubits and Toffoli gates to 2log($n$) and $n$, respectively, in the bit precision $n$. We demonstrate the algorithms using the IBM Quantum Experience cloud services. The further reduced complexity of the present algorithms brings the black-box quantum state preparation closer to reality.

## 1. Introduction

Black-box quantum state preparation is an important building block for many higher-level quantum algorithms. It is usually employed to prepare initial states as input to quantum algorithms or perform intermediate state evolutions in various scenarios, such as Hamiltonian simulation [1-3], quantum linear systems solving [4,5] and quantum machine learning [6,7]. In the sense of quantum data encoding, black-box state preparation actually implements a quantum version of digital-to-analog conversion [8].

The first black-box state preparation algorithm was given by Grover [9,10], which is an extension of his quantum searching algorithm [11]. Grover's algorithm needs to do arithmetic on quantum computer (calculating arcsines to get the rotation angles), which make the complexity rather high. Since then there has been a series of new techniques and improvements with different restrictions on the amplitudes to be loaded [12,13]. Most recently, Sanders et al. devised a modification to Grover's algorithm, which avoids the need to do arithmetic and make the black-box state preparation more practical [14]. Inspired by Sanders et al.'s work, Bausch further improved the cost through a novel design of initial and intermediate states in the amplitude amplification procedure [15]. In the present work, we show that the complexity can be reduced further using the technique of linear combination of unitaries (LCU).

Here the word "black-box" means that the amplitudes of the state are unknown a priori and are accessed through an oracle or black-box [14]. Specifically, if the amplitude vector to be loaded is $\vec{x} = (x_0, x_1, ..., x_{d-1})$ with $0 \leq x_j < 1$ for each element, then there exists an oracle $U_o$ creating the following state

$$U_o\left(\frac{1}{\sqrt{d}}\sum_{j=0}^{d-1}|j\rangle|0^n\rangle\right) = \frac{1}{\sqrt{d}}\sum_{j=0}^{d-1}|j\rangle|x_j\rangle, \qquad (1)$$

where the superscript $n$ in the second register represents the required number of qubits. That is, each amplitude $x_j$ is encoded with $n$-bit precision in the second register. It is assumed that the oracle $U_o$ can be invoked, as required, any number of times and the complexity is $O(1)$. Starting with Eq. (1), the basic task of black-box state preparation is to produce the state

$$|\text{target}\rangle = \frac{1}{\|\vec{x}\|_2}\sum_{j=0}^{d-1}x_j|j\rangle. \qquad (2)$$

In this work, we provide two algorithms for black-box state preparation based on the technique of linear combination of unitaries (LCU). Our first algorithm is based on the standard form of LCU [5,16] and has a cost of 2log($n$)-1 additional qubits and 3$n$-4 Toffoli gates (where $n$ has the same meaning as that in Eq. (1)). The second one is based on a modified version of LCU and has a cost of $n$+2 additional qubits and $n$ Toffoli gates. For comparison, the cost of Sanders et al.'s algorithm is $n$+2 additional qubits and 2$n$-1 Toffoli gates [14] and the cost of Bausch's log($n$) additional qubits and 2$n$log($n$) Toffoli gates [15]. Therefore, our algorithm improve upon Sanders et al.'s and Bausch's algorithm by reducing both the required ancilla qubits and non-Clifford gates.

This paper is organized as follows. In section 2, we discuss the standard LCU-based state preparation algorithm. The second method based on the modified LCU is discussed in section 3. In section 4, we provide the detailed complexity analysis of the two algorithms, as well as the comparison with other works. Section 5 presents the demonstration results of the two algorithms on the quantum device of IBM Quantum Experience. Finally, conclusions are given in section 6.

**2. Standard LCU-based state preparation algorithm**

We start by introducing the technique of linear combination of unitaries (LCU) [16]. LCU is used to implement any (not necessarily unitary) operator $V$ that can be written as a linear combination of unitaries $U_i$, i.e. $V = \sum_i a_i U_i$ with $a_i > 0$. Let $A$ be a unitary that maps $|0^m\rangle$ to $\frac{1}{\sqrt{a}}\sum_i \sqrt{a_i}|i\rangle$ with $a = \sum_i a_i$. Let $U$ be a block encoding of $U_i$, i.e. $U = \sum_i |i\rangle\langle i| \otimes U_i$. Then for any $n$-qubit state $|\phi\rangle$, the unitary $W = A^\dagger U A$ satisfies

$$W|0^m\rangle|\phi\rangle = \frac{1}{a}|0^m\rangle V|\phi\rangle + |\Phi^\perp\rangle, \qquad (3)$$

where the unnormalized state $|\Phi^\perp\rangle$ satisfies $(|0^m\rangle\langle 0^m| \otimes \mathbf{1})|\Phi^\perp\rangle = 0$. In other words,

the operator $V$ can be implemented through unitary $W$ in a probabilistic way, and the successful outcome occurs with probability $\left(\|V|\phi\rangle\|/a\right)^2$. With the aid of oblivious amplitude amplification [16,17], operator $V$ can be implemented (almost) exactly with $O(a/\|V|\phi\rangle\|)$ uses of $W$ and its inverses.

The idea that LCU can be utilized for black-box state preparation is based on the basic fact that an $n$-bit data $x_j \in [0,1)$ can be decomposed as $x_j = \sum_{i=0}^{n-1} \frac{1}{2^{i+1}} x_{j,n-i-1}$. The factors $1/2^{i+1}$ can be taken as the coefficients $a_i$ of the operator $V$ in LCU, and the binary $0.x_{j,n-1}x_{j,n-2}\ldots x_{j,1}x_{j,0}$ corresponding to the binary string in the second register in Eq. (1) are prepared by the unitaries $U_i$.

In our LCU-based black-box state preparation algorithms, the most important trick is the way of performing the corresponding transform $A$, which takes the state $|0^m\rangle$ to

$$|\psi_A\rangle = \frac{1}{\sqrt{a}} \sum_{i=0}^{n-1} \frac{1}{\sqrt{2^{i+1}}} |i\rangle$$

with $a = (2^n - 1)/2^n$ being the normalization constant. It is interesting to note that the state $|\psi_A\rangle$ is unentangled, and can be factorized as a tensor product of single-qubit state, i.e. $|\psi_A\rangle = \bigotimes_{i=0}^{m-1}\left(\cos\theta_i|0\rangle + \sin\theta_i|1\rangle\right)$ with $\theta_i \in (0, \pi/4)$. Details about the factorization are shown in Appendix A. As shown in Fig.1, we can create the state $|\psi_A\rangle$ through $\log(n)$ single-qubit $R_y$ gates.

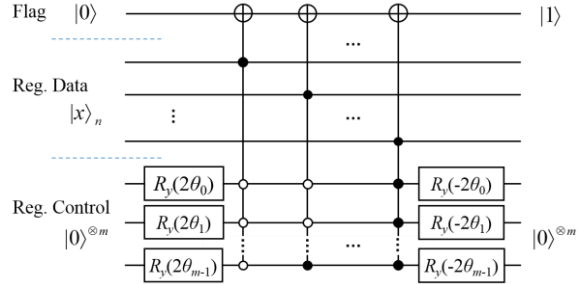

Fig. 1 The circuit for performing the map $|x\rangle \to x|x\rangle$ using the standard form of LCU algorithm. The circuit consists of three registers, i.e. the Flag, Data and Control register with the number of qubits of one, $n$ and $m$, respectively, and $m = \log(n)$. For the cases of $m = \lceil \log n \rceil$, the circuits can be obtained by extending the above result directly.

The rotation angles of $R_y$ gates in Fig.1 is determined in the following way. More details about the derivation of the rotation angles can be found in Appendix A. For each $R_y$ gates, the angle $\theta_i$ is determined by $\cos\theta_i = \sqrt{\dfrac{2^{2^{m-1-i}}}{2^{2^{m-1-i}}+1}}$, and it creates the state $\sqrt{\dfrac{2^{2^{m-1-i}}}{2^{2^{m-1-i}}+1}}|0\rangle + \sqrt{\dfrac{1}{2^{2^{m-1-i}}+1}}|1\rangle$. For example, the first angle $\theta_0$ satisfies $\cos\theta_0 = \sqrt{\dfrac{2^{2^{m-1}}}{2^{2^{m-1}}+1}}$, and it creates the state $\sqrt{\dfrac{2^{2^{m-1}}}{2^{2^{m-1}}+1}}|0\rangle + \sqrt{\dfrac{1}{2^{2^{m-1}}+1}}|1\rangle$; the second angle $\theta_1$ satisfies $\cos\theta_1 = \sqrt{\dfrac{2^{2^{m-2}}}{2^{2^{m-2}}+1}}$, and it creates the state $\sqrt{\dfrac{2^{2^{m-2}}}{2^{2^{m-2}}+1}}|0\rangle + \sqrt{\dfrac{1}{2^{2^{m-2}}+1}}|1\rangle$; and the last angle $\theta_{m-1}$ satisfies $\cos\theta_{m-1} = \sqrt{\dfrac{2}{3}}$, and it creates the state $\sqrt{\dfrac{2}{3}}|0\rangle + \sqrt{\dfrac{1}{3}}|1\rangle$. So the $m$ $R_y$ gates can produce the state $|\psi_A\rangle$ in the Control register. In practice, when implementing the $R_y$ gates, the angles $\theta_i$ would have truncation errors, which will finally result into the error of the target state in Eq. (2). In appendix B, we show that in order to guarantee that the amplitudes $x_j$ are to be loaded with $n$-bit precision, the truncation error of the rotation angles should be less than about $2^{-n-5}$.

Following the group of $R_y$ rotations, namely the unitary $A$, is the $(m+1)$-controlled NOT operations as shown in Fig.1. It makes the phases $x_{j,n-i-1}/\sqrt{2^{i+1}}$ kick-back onto the Flag qubit. Finally the inverse operation of unitary $A$ is performed. The overall state evolution through the circuit in Fig. 1 is as follows,

$$\begin{aligned}&\frac{1}{\sqrt{d}}\sum_{j=0}^{d-1}|j\rangle|x_j\rangle_{\text{Data}}|0^m\rangle_{\text{Control}}|0\rangle_{\text{Flag}}\\&\mapsto \frac{\|\vec{x}\|_2}{a\cdot\sqrt{d}}\sum_{j=0}^{d-1}\frac{x_j}{\|\vec{x}\|_2}|j\rangle|x_j\rangle_{\text{Data}}|0^m\rangle_{\text{Control}}|1\rangle_{\text{Flag}}+|\omega\rangle_\perp\end{aligned}, \quad (4)$$

where $a$ is the normalization constant as mentioned before and $|\omega\rangle_\perp$ is the superposition state orthogonal to $|0^m\rangle_{\text{Control}}|1\rangle_{\text{Flag}}$. If we measure the Control and Flag register and obtain a result $|0^m\rangle|1\rangle$, then the target state is created successfully with a success probability of $(\|\vec{x}\|_2/a\sqrt{d})^2$. Note that the constant $a$ is less than one, so it can increase the success probability. Moreover, by performing $O(a\sqrt{d}/\|\vec{x}\|_2)$ rounds of amplitude amplification procedure [17], the success probability can increase nearly to one. At last, the Data register need to be reset by applying the oracle $U_o$ once more.

## 3. Modified LCU-based state preparation algorithm

In the above standard LCU-based algorithm, multi-controlled NOT operations are required which is essentially due to the way of encoding the combination coefficients $a_i$. For the state $|\psi_A\rangle = \frac{1}{\sqrt{a}}\sum_{i=0}^{n-1}\frac{1}{\sqrt{2^{i+1}}}|i\rangle$, the $n$ phase factors $1/\sqrt{2^{i+1}}$ are indexed by log($n$)-bit binary. This is indeed the reason that the number of additional qubit in our standard LCU-based algorithm is exponential lower than that of Sanders et al.'s method. However, in practice the multi-controlled operations could not be implemented efficiently [18]. Additionally, the $R_y$ gates need to be implemented with enough precision, which would not be easy in practice. Therefore, we propose an alternative black-box state preparation algorithm using a modified form of LCU.

The main idea is straightforward. If the $n$ phase factors were indexed by a qudit with $n$ levels, then the multi-controlled operation could be avoided. In the scenario of qubit, the $n$ phase factors should be indexed by a serial of binary like (1000…0), (0100…0) and (0010…0) and so on. That is, in this case the operation $A$ should create such a state

$$|\psi_{A'}\rangle = \frac{1}{\sqrt{a}}\sum_{i=0}^{n-1}\frac{1}{\sqrt{2^{i+1}}}|2^{n-1-i}\rangle$$

where $a$ is the normalization constant. Note that this state is similar as the amplitude gradient state in Ref. [15]. Such a state can be created easily by applying a cascade of controlled-$H$ and CNOT gates as shown in Fig. 2.

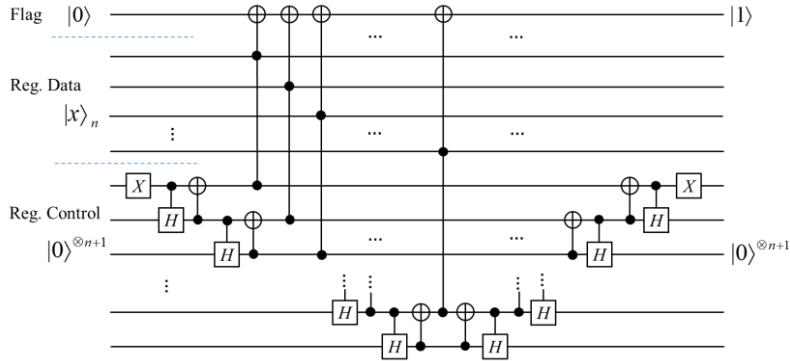

Fig. 2 The circuit for performing the map $|x\rangle \rightarrow x|x\rangle$ using a modified form of LCU algorithm. The circuit consists of three registers, i.e. the Flag, Data and Control register with the number of qubits of one, $n$ and $n+1$, respectively.

Let us now sketch the state evolution through the circuit in Fig. 2. In the Control register, after the first group of C-$H$ and CNOT gates the state turns to $1/\sqrt{2}|10\rangle + 1/\sqrt{2}|01\rangle$; after the second group of C-$H$ and CNOT gates, the state is $\frac{1}{\sqrt{2}}|100\rangle + \frac{1}{\sqrt{2^2}}|010\rangle + \frac{1}{\sqrt{2^2}}|001\rangle$. Thus, the final control state is

$$\frac{1}{\sqrt{2}}|10...0\rangle + \frac{1}{\sqrt{2^2}}|010...0\rangle + \cdots + \frac{1}{\sqrt{2^n}}|0...010\rangle + \frac{1}{\sqrt{2^n}}|0...01\rangle = \sum_{i=0}^{n-1}\frac{1}{\sqrt{2^{i+1}}}|2^{n-i}\rangle + \frac{1}{\sqrt{2^n}}|1\rangle. \quad (5)$$

Note that this state is slightly different with the state $|\psi_{A'}\rangle$ mentioned above. This state contains $n+1$ qubits, and the last superposition term $|1\rangle$ is left unused as shown in Fig. 2. Here we remark that by applying controlled $R_y$ gates instead of the C-H gates, the above technique would be taken as a general high-efficiency method for implementing operation $A$ in LCU algorithm.

The overall state evolution through the circuit in Fig. 2 is similar as before,

$$\frac{1}{\sqrt{d}}\sum_{j=0}^{d-1}|j\rangle|x_j\rangle_{\text{Data}}|0^{n+1}\rangle_{\text{Control}}|0\rangle_{\text{Flag}}$$
$$\mapsto \frac{\|\vec{x}\|_2}{\sqrt{d}}\sum_{j=0}^{d-1}\frac{x_j}{\|\vec{x}\|_2}|j\rangle|x_j\rangle_{\text{Data}}|0^{n+1}\rangle_{\text{Control}}|1\rangle_{\text{Flag}}+|\omega\rangle_{\perp}. \qquad (6)$$

As before, if we measure the Control and Flag register and obtain a result $|0^{n+1}\rangle|1\rangle$, then the target state is created successfully. By performing $O(\sqrt{d}/\|\vec{x}\|_2)$ rounds of amplitude amplification, the success probability can increase nearly to one.

## 4. Complexity analysis

Now we analyze the cost of our two algorithms and make a comparison with that of the so far best results as we know. For the standard LCU-based algorithm, the major contributor to the complexity is the $n$ multi-controlled NOT gates as shown in Fig. 1. In general, one $(m+1)$-controlled NOT gates can be decomposed into $2m$-1 Toffoli gates by using $m$-1 ancilla qubits [19]. However, after decomposition of all the $n$ $(m+1)$-controlled NOT gates ($m=\log(n)$), there appears many overlaps among them, which finally merge into $(3n-4)$ Toffoli gates and $(n-2)$ CNOT gates. This phenomena is similar to the unary iteration idea of ref. [20].

For the modified LCU-based algorithm, the complexity is obvious as shown in Fig. 2. The number of ancilla qubits is $n+2$ and Toffoli gates is $n$. The total cost of our two algorithms, as well as that of Sanders et al.'s [14] and Bausch's [15] algorithm, is summarized in Table 1.

For the Sanders et al.'s algorithm, the major cost results from the comparison operation performed on the computational basis. They prescribe two choices to perform the comparison operation, which have complementary advantage in the cost of additional qubits and non-Clifford gates. The first choice [21] results into a complexity of $2n+1$ in additional qubits and $n$ in Toffoli gates, and the second choice [22] a complexity of $n+2$ ancillas and $2n-1$ Toffoli gates. Note that for the first choice, the authors of ref. [14] did not take the temporary logical-AND operations into account. In fact, two Toffoli gates appearing in compute/uncompute pairs can be replaced by a logical-AND pair if the intermediate operations are not sensitive to phase errors of the control qubits of the Toffoli gates. Thus, if the temporary logical-AND operation is applied in our standard LCU-based algorithm, one $(m+1)$-controlled NOT gate can be implemented by $m$-1 temporary logical-AND pairs and 1 Toffoli gate. Then the cost of Toffoli gates in our standard LCU-based algorithm is reduced to $n$. To avoid the confusion, Table 1 only presents Sanders et al.'s results based on the second choice.

Being different with Sanders et al.'s and our algorithms, Bausch's algorithm has no distinguishable process of amplitude transduction. In order to make a reasonable comparison, we list the total cost of the two fundamental building blocks of the algorithm in the table. Specifically, the cost of phase kickback oracle is $\log(n)$ in ancillary qubits and $2n\log(n)$ in Toffoli gates. The $\sqrt{\text{SWAP}}$ gates are used to create the amplitude gradient state. While as shown by the author, in many cases further optimizations can be applied considering the causality cone of the Z-gate, in the table we just present the general complexity expressions as a comparison.

Tab. 1 Cost of the four algorithms for amplitude transduction as $|x\rangle \to x|x\rangle$

|              | Additional qubit | Toffoli     | $\sqrt{\text{SWAP}}$ | CNOT   |
| ------------ | ---------------- | ----------- | -------------------- | ------ |
| Sanders+ [14]| $n+2$            | $2n-1$      |                      | $4n-3$ |
| Bausch [15]  | $\log(n)$        | $2n\log(n)$ | $n$                  |        |
| Standard-LCU | $2\log(n)-1$     | $3n-4$      |                      | $n-2$  |
| Modified-LCU | $n+2$            | $n$         |                      | $4n$   |

The four algorithms listed in Table 1 will be further embedded into the amplitude amplification process to increase the success probability. The cost shown in Table 1 is only for performing the amplitude transduction as $|x\rangle \to x|x\rangle$. Generally, the amplitude amplification consists of several Grover iterations. Each iteration contains two reflections, that is, first reflects about the target-vertical states and then about the initial states. Here we remark that the reflection transform need to perform the multi-controlled $Z$ operations, which would cost as much as the transduction process. In Table 2, we summarize the cost of implementing the two kinds of reflections. In the table, one $k$-controlled NOT gate is taken to be decomposed into $2k-3$ Toffoli gates by using $(k-2)$ ancilla qubits. Note that the "ancillas" in the table represent the extra ancilla qubits with respect to the "Additional qubits" in Table 1. In practice, the new techniques of single-step implementation of multi-qubit controlled gates [23] will be great useful for executing black-box state preparation algorithm.

Tab. 2 Cost of the two reflections in each amplitude amplification iteration

|             | Sanders+ | Bausch         | Standard-LCU   | Modified-LCU |
| ----------- | -------- | -------------- | -------------- | ------------ |
| Ancillas    | $2n-3$   | $n+\log(n)-3$  | $n$            | $2n-1$       |
| Toffoli [a] | $2n-3$   | $2\log(n)-5$   | $2\log(n)-3$   | $2n-1$       |
| Toffoli [b] | $4n-3$   | $2n+2\log(n)-5$| $2n+2\log(n)-3$| $4n-1$       |

[a] For the reflections about the target-vertical states, namely the Grover oracle $O_f$
[b] For the reflections about the initial states, namely the Grover diffusion operator

## 5. Demonstration of the algorithms

We demonstrate our two algorithms using the IBM Quantum Experience cloud services [24] by preparing a state $|\varphi\rangle = 5^{-1/2}(|0\rangle + 2|1\rangle)$. The overall structure of the two circuits is illustrated in Fig. 3. In the standard LCU-based algorithm, the number of qubits and elementary gates is 8 and 57 respectively; in the modified LCU-based algorithm, it is 12 and 91 respectively. For the two demonstration circuits, one round of amplification is enough, and the success probability is in theory 99.11% and 88.13% respectively. The ibmq_qasm_simulator quantum computer is used, and the circuits were executed 10 times with 8192 shots in each time. The averaged successful probability is 99.01% and 88.04% (accessed on 2021.04) respectively. The original QASM codes for the demonstration circuits can be found in the supplementary material of this paper.

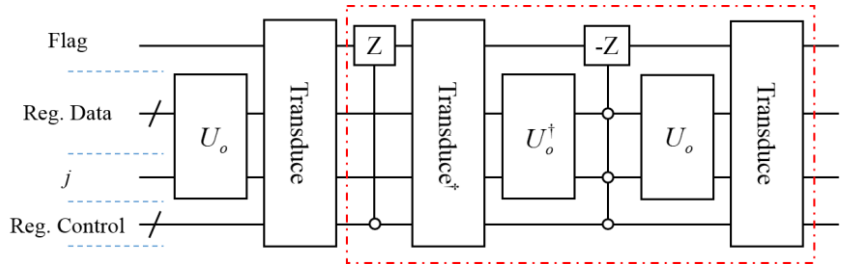

Fig. 3 The overall circuit used to demonstrate our two algorithms for state preparation. The $U_o$ is the oracle encoding the data into quantum computer. The Transduce module represents the standard (modified) LCU-based algorithm. The dotted box denotes one round of amplitude amplification.

## 6. Conclusions

In the present work, we develop two black-box quantum state preparation algorithms based on the linear combination of unitaries (LCU) algorithm. The main difference of the two algorithms is the way of performing the transform $A$ in LCU. The standard LCU-based algorithm has a complexity of $2\log(n)-1$ in additional qubits and $3n-4$ in Toffoli gates, and the modified LCU-based algorithm a complexity of $n+2$ in additional qubits and $n$ in Toffoli gates. It is interesting to note that our modified scheme provides an efficient way to implement LCU algorithm without the subroutine of black-box quantum state preparation for implementing unitary $A$ as before. The present algorithms are demonstrated using the IBM Quantum Experience.

Now the complexity of black-box quantum state preparation is very low, which is nearly linear to the precision of the data with logarithmic additional qubits. For the next step, we will study the practical executability of the algorithms in the scenario of NISQ (near-term intermediate-scale quantum devices). It is intriguing to know to what extent the noise would affect the outputs of the algorithm and how to reduce the error.

**Acknowledgements**

We acknowledge the use of IBM Quantum services for this work. The views expressed are those of the authors, and do not reflect the official policy or position of IBM or the IBM Quantum team. The present work is supported by the National Natural Science Foundation of China (Grant No. 12005212, 61575180, 61701464) and the Pilot National Laboratory for Marine Science and Technology (Qingdao).

**Appendix A**

In this appendix, we give the detailed derivation of factorizing the state $|\psi_A\rangle$ used in the standard LCU-based quantum state preparation algorithm. The basic idea is to break the serial of coefficients in half step-by-step as follows,

$$
\begin{aligned}
|\psi_A\rangle &= \sqrt{\frac{2^n}{2^n-1}} \sum_{i=0}^{n-1} \frac{1}{\sqrt{2^{i+1}}} |i\rangle \\
&= \sqrt{\frac{2^{2^m}}{2^{2^m}-1}} \left( \sum_{i=0}^{2^{m-1}-1} \frac{1}{\sqrt{2^{i+1}}} |i\rangle_m + \sum_{j=2^{m-1}}^{2^m-1} \frac{1}{\sqrt{2^{j+1}}} |j\rangle_m \right) \\
&= \sqrt{\frac{2^{2^m}}{2^{2^m}-1}} \left( \sum_{k=0}^{2^{m-1}-1} \frac{1}{\sqrt{2^{k+1}}} |0\rangle|k\rangle_{m-1} + \sum_{k=0}^{2^{m-1}-1} \frac{1}{\sqrt{2^{k+1+2^{m-1}}}} |1\rangle|k\rangle_{m-1} \right) \\
&= \sqrt{\frac{2^{2^m}}{2^{2^m}-1}} \left( |0\rangle + \frac{1}{\sqrt{2^{2^{m-1}}}} |1\rangle \right) \otimes \sum_{k=0}^{2^{m-1}-1} \frac{1}{\sqrt{2^{k+1}}} |k\rangle_{m-1} \\
&= \cdots \\
&= \sqrt{\frac{2^{2^m}}{2^{2^m}-1}} \bigotimes_{i=0}^{m-2} \left( |0\rangle + \frac{1}{\sqrt{2^{2^{m-1-i}}}} |1\rangle \right) \otimes \sum_{k=0}^{1} \frac{1}{\sqrt{2^{k+1}}} |k\rangle_1 \\
&= \sqrt{\frac{2^{2^m}}{2^{2^m}-1}} \bigotimes_{i=0}^{m-2} \left( |0\rangle + \frac{1}{\sqrt{2^{2^{m-1-i}}}} |1\rangle \right) \otimes \left( \frac{1}{\sqrt{2}} |0\rangle + \frac{1}{2} |1\rangle \right)
\end{aligned} \quad (A1)
$$

Next, the first factor $\sqrt{2^{2^m}/2^{2^m}-1}$ can be factorized as follows,

$$
\begin{aligned}
\sqrt{\frac{2^{2^m}}{2^{2^m}-1}} &= \sqrt{\frac{2^{2^{m-1}} 2^{2^{m-1}}}{(2^{2^{m-1}}+1)(2^{2^{m-1}}-1)}} \\
&= \sqrt{\frac{2^{2^{m-1}}}{2^{2^{m-1}}+1}} \sqrt{\frac{2^{2^{m-1}}}{2^{2^{m-1}}-1}} \\
&= \sqrt{\frac{2^{2^{m-1}}}{2^{2^{m-1}}+1}} \sqrt{\frac{2^{2^{m-2}}}{2^{2^{m-2}}+1}} \sqrt{\frac{2^{2^{m-2}}}{2^{2^{m-2}}-1}} \\
&= \cdots \\
&= \left( \prod_{i=0}^{m-2} \sqrt{\frac{2^{2^{m-1-i}}}{2^{2^{m-1-i}}+1}} \right) \cdot \sqrt{\frac{4}{3}}
\end{aligned} \quad (A2)
$$

Associate each factor of the above equation with the corresponding single-qubit states of Eq. (A1), then we have

$$|\psi_A\rangle = \bigotimes_{i=0}^{m-2}\left(\sqrt{\frac{2^{2^{m-1-i}}}{2^{2^{m-1-i}}+1}}|0\rangle + \sqrt{\frac{1}{2^{2^{m-1-i}}+1}}|1\rangle\right) \otimes \left(\sqrt{\frac{2}{3}}|0\rangle + \sqrt{\frac{1}{3}}|1\rangle\right). \quad (A3)$$

Thus, the state $|\psi_A\rangle$ is actual a product state and can be expressed as $|\psi_A\rangle = \bigotimes_{i=0}^{m-1}\left(\cos\theta_i|0\rangle + \sin\theta_i|1\rangle\right)$ with $\theta_i \in (0, \pi/4)$. This equation can be implemented easily using a group of single-qubit $R_y$ rotations.

**Appendix B**

In practical implementation of the $R_y$ rotations in the standard LCU-based algorithm, the rotation angles $\theta_i$ would have truncation errors. The error will reduce the fidelity of the created state. In this appendix, we will analyze the relation between the precision of the output and the truncation error of $R_y$ gates.

First, we argue that the truncation error of the angles will lead to the maximum error in the amplitude before the basis $|0\rangle$ and $|2^m-1\rangle$ in the control state

$$|\psi_A\rangle = \frac{1}{\sqrt{a}}\sum_{i=0}^{n-1}\frac{1}{\sqrt{2^{i+1}}}|i\rangle = \bigotimes_{i=0}^{m-1}\left(\cos\theta_i|0\rangle + \sin\theta_i|1\rangle\right) \quad (B1)$$

with $\theta_i \in (0, \pi/4)$. In the interval $(0, \pi/4)$, we have the following facts,

$$\theta_i' < \theta_i; \quad \cos\theta_i' > \cos\theta_i; \quad \sin\theta' < \sin\theta_i, \quad (B2)$$

where $\theta_i'$ is the truncated angle and $\theta_i$ is the real angle. That is, all the truncated cosine values become larger and the truncated sine values become smaller. Thus, the maximum accumulation error should happen for $\prod_{i=0}^{m-1}\cos\theta_i$ and $\prod_{i=0}^{m-1}\sin\theta_i$, which is the amplitude of the basis $|0\rangle$ and $|2^m-1\rangle$, respectively. In the following, we will calculate the errors of the two significant cases.

Assume $\theta_i = \omega_i\pi$ with $\omega_i \in (0, 1/4)$, and the truncation error of the angular coefficient $\omega_i$ is expressed as $\Delta_i = \omega_i - \omega_i' \leq 2^{-k}$ with $k$ large enough. Then the error of the amplitude of basis $|0\rangle$ is obtained as follows,

$$\begin{aligned}
\varepsilon_{|0\rangle} &= \prod_{i=0}^{m-1}\cos\omega_i'\pi - \prod_{i=0}^{m-1}\cos\omega_i\pi \\
&= \cos(\omega_{m-1}-\Delta_{m-1})\pi\prod_{i=0}^{m-2}\cos\omega_i'\pi - \cos\omega_{m-1}\pi\prod_{i=0}^{m-2}\cos\omega_i\pi \\
&= \begin{pmatrix}\cos\omega_{m-1}\pi\cos\Delta_{m-1}\pi + \\ \sin\omega_{m-1}\pi\sin\Delta_{m-1}\pi\end{pmatrix}\prod_{i=0}^{m-2}\cos\omega_i'\pi - \cos\omega_{m-1}\pi\prod_{i=0}^{m-2}\cos\omega_i\pi \\
&\approx \begin{bmatrix}\cos\omega_{m-1}\pi\cdot(1-(\Delta_{m-1}\pi)^2/2) \\ +\sin\omega_{m-1}\pi\cdot(\Delta_{m-1}\pi)\end{bmatrix}\prod_{i=0}^{m-2}\cos\omega_i'\pi - \cos\omega_{m-1}\pi\prod_{i=0}^{m-2}\cos\omega_i\pi \\
&< \cos\omega_{m-1}\pi\left[\prod_{i=0}^{m-2}\cos\omega_i'\pi - \prod_{i=0}^{m-2}\cos\omega_i\pi\right] + 2^{-k+2} \\
&< \cos\omega_{m-1}\pi\left[\cos\omega_{m-2}\pi\left(\prod_{i=0}^{m-3}\cos\omega_i'\pi - \prod_{i=0}^{m-3}\cos\omega_i\pi\right) + 2^{-k+2}\right] + 2^{-k+2} \\
&= \cdots \\
&< \cos\omega_{m-1}\pi\left[\cos\omega_{m-2}\pi\left(\cdots\left(\cos\omega_1\pi\left(2^{-k+2}\right) + 2^{-k+2}\right)\cdots\right) + 2^{-k+2}\right] + 2^{-k+2} \\
&< \cos\omega_{m-1}\pi\left[\cos\omega_{m-2}\pi\left(\cdots\left(2\cdot 2^{-k+2}\right)\cdots\right) + 2^{-k+2}\right] + 2^{-k+2} \\
&= \cdots \\
&< \cos\omega_{m-1}\pi\cdot(m-1)\cdot 2^{-k+2} + 2^{-k+2} \\
&< m\cdot 2^{-k+2} \\
&< 2^{-k+2+\lceil\log m\rceil}
\end{aligned} \quad \text{(B3)}$$

The error of the amplitude of basis $|2^m-1\rangle$ is obtained as follows,

$$\begin{aligned}
\varepsilon_{|2^m-1\rangle} &= \prod_{i=0}^{m-1}\sin\omega_i\pi - \prod_{i=0}^{m-1}\sin\omega_i'\pi \\
&= \sin\omega_{m-1}\pi\prod_{i=0}^{m-2}\sin\omega_i\pi - \sin(\omega_{m-1}-\Delta_{m-1})\pi\prod_{i=0}^{m-2}\sin\omega_i'\pi \\
&\approx \sin\omega_{m-1}\pi\prod_{i=0}^{m-2}\sin\omega_i\pi - \begin{pmatrix}\sin\omega_{m-1}\pi\cdot(1-(\Delta_{m-1}\pi)^2/2) \\ -\cos\omega_{m-1}\pi\cdot(\Delta_{m-1}\pi)\end{pmatrix}\prod_{i=0}^{m-2}\sin\omega_i'\pi \\
&< \sin\omega_{m-1}\pi\left(\prod_{i=0}^{m-2}\sin\omega_i\pi - \prod_{i=0}^{m-2}\sin\omega_i'\pi\right) + 2^{-k+2-(m-1)} \\
&< \sin\omega_{m-1}\pi\left[\sin\omega_{m-2}\pi\left(\prod_{i=0}^{m-3}\sin\omega_i\pi - \prod_{i=0}^{m-3}\sin\omega_i'\pi\right) + 2^{-k+2-(m-2)}\right] + 2^{-k+2-(m-1)} \\
&= \cdots \\
&< \sin\omega_{m-1}\pi\cdot 2^{-k+2-(m-3)} + 2^{-k+2-(m-1)} \\
&< 2^{-k+2-(m-2)}
\end{aligned} \quad \text{(B4)}$$

As shown by Eq. (B3) and (B4), the maximum error occurs in the amplitude of basis $|0\rangle$. So the amplitude error of basis $|0\rangle$, $\varepsilon_{|0\rangle}$ is taken as the upper bound of the amplitude error of all basis of state $|\psi_A\rangle$.

The goal is to transduce $x_j = \sum_{i=0}^{n-1} \frac{1}{2^{i+1}} x_{j,n-i-1}$ as the amplitude in the target state, while the above error $\varepsilon_{|0\rangle}$ is the error of the factor $1/\sqrt{2^{i+1}}$. So next, we calculate the square of the error $\varepsilon_{|0\rangle}$ and sum over to get the final error. Let $\sqrt{a_i} = 1/\sqrt{2^{i+1}}$, then the upper bound of the error of each $a_i$ can be obtained as

$$|a_i - a_i'| = \left|\sqrt{a_i} + \sqrt{a'}\right| \cdot \left|\sqrt{a_i} - \sqrt{a'}\right| < \left|2\sqrt{a_i} + \varepsilon_{|0\rangle}\right| \cdot \varepsilon_{|0\rangle} \approx 2^{-k+3+\lceil \log m \rceil - (i+1)/2}. \quad (B5)$$

Sum over the $n$ terms, then we have

$$\varepsilon_x = \sum_{i=0}^{n-1} |a_i - a_i'| < 2^{-k+3+\lceil \log m \rceil} \sum_{i=0}^{n-1} 2^{-(i+1)/2} = 2^{-k+3+\lceil \log m \rceil} \frac{1 - 1/\sqrt{2^n}}{\sqrt{2} - 1} < 2^{-k+5+\lceil \log m \rceil}. \quad (B6)$$

In one word, if we want to transduce a data $x$ as the amplitude with $n$-bit precision, then the rotation angles of the $R_y$ gates in the standard LCU-based algorithm should have at least $n + 5 + \lceil \log \lceil \log n \rceil \rceil$-bit precision.

**References**
[1] Childs A M 2010 On the relationship between continuous-and discrete-time quantum walk *Commun. Math. Phys.* **294** 581
[2] Berry D W, Childs A M, Cleve R, Kothari R and Somma R D 2015 Simulating Hamiltonian dynamics with a truncated Taylor series *Phys. Rev. Lett.* **114** 090502
[3] Low G H and Chuang I L 2017 Optimal Hamiltonian simulation by quantum signal processing *Phys. Rev. Lett.* **118** 010501
[4] Harrow A W, Hassidim A and Lloyd S 2009 Quantum algorithm for linear systems of equations *Phys. Rev. Lett.* **103** 150502
[5] Childs A M, Kothari R and Somma R D 2017 Quantum algorithm for systems of linear equations with exponentially improved dependence on precision. *SIAM Journal on Computing* **46** 1920 (*arXiv*: 1511.02306)
[6] Wiebe N, Braun D and Lloyd S 2012 Quantum algorithm for data fitting *Phys. Rev. Lett.* **109** 050505
[7] Kerenidis I and Prakash A 2020 Quantum gradient descent for linear systems and least squares *Phys. Rev. A* **101** 022316
[8] Mitarai K, Kitagawa M and Fujii K 2019 Quantum analog-digital conversion *Phys. Rev. A* **99** 012301
[9] Grover L K 2000 Synthesis of Quantum Superpositions by Quantum Computation *Phys. Rev. Lett.* **85** 1334
[10] Grover L K and Rudolph T 2002 Creating superpositions that correspond to efficiently

integrable probability distributions *arXiv*: quant-ph/0208112

[11] Grover L K 1997 Quantum mechanics helps in searching for a needle in a haystack *Phys. Rev. Lett.* **79** 325

[12] Soklakov A and Schack R 2006 Efficient state preparation for a register of quantum bits *Phys. Rev. A* **73**, 012307

[13] Zoufal C, Lucchi A and Woerner S 2019 Quantum Generative Adversarial Networks for learning and loading random distributions *npj Quantum Inf.* **5** 103

[14] Sanders Y R, Low G H, Scherer A and Berry D W 2019 Black-Box Quantum State Preparation without Arithmetic *Phys. Rev. Lett.* **122** 020502

[15] Bausch J 2009 Fast Black-Box Quantum State Preparation *arXiv*: 2009.10709

[16] Kothari R 2014 Efficient algorithms in quantum query complexity, PhD thesis *University of Waterloo* Chap 2

[17] Brassard G, Høyer P, Mosca M and Tapp A 2000 Quantum Amplitude Amplification and Estimation *arXiv*: quant-ph/0005055

[18] Barenco A, Bennett C H, Cleve R, DiVincenzo D P, Margolus N, Shor P, Sleator T, Smolin J A and Weinfurter H 1995 Elementary gates for quantum computation *Phys. Rev. A* **52**, 3457

[19] Nielsen M A and Chuang I L 2010 Quantum computation and quantum information. *Cambridge University Press*, *Cambridge* Chap 4

[20] Babbush R, Gidney C, Berry D W, Wiebe N, McClean J, Paler A, Fowler A and Neven H 2018 Encoding electronic spectra in quantum circuits with linear T complexity *Phys. Rev. X* **8**, 041015

[21] Gidney C 2018 Halving the cost of quantum addition Quantum **2** 74

[22] Cuccaro S A, Draper T G, Kutin S A and Moulton D P 2004 A new quantum ripple-carry addition circuit *arXiv*: quant-ph/0410184

[23] Rasmussen S E, Groenland K, Gerritsma R, Schoutens K and Zinner N T 2020 Single-step implementation of high-fidelity n-bit Toffoli gates *Phys. Rev. A* **101**, 022308

[24] *ibmq_qasm_simulator* v0.1.547 *IBM Quantum team*. https://quantum-computing.ibm.com (accessed on 2021.04)